\documentclass[aps,prl,twocolumn,showpacs]{revtex4}
\usepackage{amsmath}
\usepackage{bm}
\usepackage{graphicx}

\begin{document}

\title{Friedel phase discontinuity and bound states in the continuum in
quantum dot systems}
\author{B. Sol\'is}
\author{M. L. Ladr\'on de Guevara}
\author{P. A. Orellana}
\affiliation{Departamento de F\'{\i}sica, Universidad Cat\'{o}lica del Norte, Casilla
1280, Antofagasta, Chile}

\begin{abstract}
In this article we study the Friedel phase of the electron transport in two
different systems of quantum dots which exhibit bound states in the
continuum (BIC). The Friedel phase jumps abruptly in the energies of the
BICs, which is associated to the vanishing width of these states, as shown
by Friedrich and Wintgen in Phys. Rev. A \textbf{31}, 3964 (1985). This odd
behavior of the Friedel phase has consequences in the charge through the
Friedel sum rule. Namely, if the energy of the BIC drops under the Fermi
energy the charge changes abruptly in a unity. We show that this behavior
closely relates with discontinuities in the conductance predicted for
interacting quantum dot systems.
\end{abstract}

\date{\today}
\maketitle

In a early work, von Neumann and Wigner \cite{vonneumann} showed that for
certain local potentials,
the Schr\"{o}dinger equation has exact solutions with energy eigenvalues
above the continuum threshold. These potentials can be constructed in one
dimension by a method suggested in the same article. Much later, several
theoretical and experimental works show the existence of these ``bound
states in the continuum'' (BICs) in different contexts. In Ref. \cite{newton}
it is shown that these states can occur in a system of coupled square well
potentials for appropriate values of the well depths and coupling strengths.
Stillinger and Herrick corrected and extended von Neumann work to consider
systems in two dimensions \cite{stillinger}, constructing some potentials
leading to BICs. Friedrich and Wintgen demonstrated
that BICs occur in natural way, not only for specific potentials,
when two resonances associated with different channels interfere
\cite{friedrich}.
In transport through mesoscopic and nanoscopic
systems, there are theoretical works showing the formation of
these states in a four-terminal
junction \cite{schult}, in a ballistic channel with intersections \cite%
{zhen-li}, and more recently, bound states in the continuum have been
theoretically discussed for systems of quantum dots \cite%
{ghost,rotter,ordonez,triple}. The existence of BICs in open quantum
billiards with variable shape is discussed in Ref. \cite{sadreev}, and in a
single-level Fano-Anderson model with a colored interaction with the
continuum in Ref. \cite{longhi}. An experimental evidence of BICs was
reported by Capasso \emph{et al.} \cite{capasso} in semiconductor
heterostructures grown by molecular beam epitaxy.

The density of states (DOS) is of great relevance to understand
transport phenomena. The DOS is related to the scattering matrix
$S$ via the Friedel sum rule (FSR), which can be stated as
\begin{equation}
\frac{d\theta_{F}}{d\varepsilon}=\pi \rho(\varepsilon) \label{fsr}
\end{equation}
where the Friedel phase $\theta_F$ is defined as
\begin{equation}
\theta_F=\sum_{l=1}^2\xi_l= \frac{1}{2i}\log{(\det{S})},
\label{friedelphase}
\end{equation}
where the phase shifts $\xi_l$ are obtained from the eigenvalues
of the scattering matrix $S$, $\lambda_l=e^{2i\xi_l}$
\cite{friedel,taniguchi}. The FSR has been central in the
understanding of the behavior of impurities in metals
\cite{langreth}, but recently has received considerable attention
in the context of low dimensional systems
\cite{lee,yeyati,taniguchi,bando,rontani,deo}. Relations between
the Friedel phase and properties such as resistance \cite{datta},
persistent current \cite{akkermans}, and capacitance \cite{gopar}
have been studied.

As emphasized by Lee \cite{lee} and Taniguchi and B\"uttiker
\cite{taniguchi}, it is not correct to identify the Friedel phase
(\ref{friedelphase}) with the phase of the amplitude of
transmission. Ref. \cite{taniguchi} studies the connection between
the Friedel phase and the transmission phase in scattering systems
in mesoscopic physics. It is shows that the transmission phase can
change abruptly in $\pi$ while the Friedel phase remains
continuous as a function of the energy. The transmission phase
discontinuity is characteristic of single electron transport and
accompanies the vanishing of the transmission amplitude
\cite{xu,kim}.

The aim of this work is to bring into discussion the behavior of
the Friedel phase in the presence of BICs in the context of
electronic transport through quantum dots. By BIC we mean a
resonance that, as consequence of quantum interference, becomes
infinitely narrow \cite{friedrich}. Since these states are
characterized by delta-shaped densities of states, the Friedel
phase, according to Eq. (\ref{friedelphase}), is discontinuous as
a function of the energy. We present results of two particular
systems: a parallel-coupled double quantum dot molecule
\cite{ghost} and two quantum dots side-attached to a quantum wire,
and we discuss the consequences that such a discontinuity has on
the conductance in presence of electron-electron interactions. We
connect our discussion with results predicted in multilevel
quantum dots in Kondo regime \cite{busser}.

\begin{figure}[t]
\includegraphics[width=7.6cm]{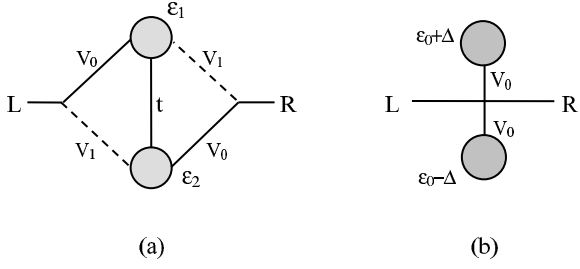}\newline
\caption{a) Double quantum dot molecule coupled to leads. b) Two
quantum dots side-coupled to a quantum wire.} \label{Fig1}
\end{figure}

The studied systems are shown schematically in Fig. \ref{Fig1}. In
both cases we consider single-level quantum dots, and we assume
equilibrium and zero temperature. The systems are modeled by
Anderson Hamiltonians. In a first stage we neglect the
electron-electron interaction. For the double quantum dot molecule of Fig. %
\ref{Fig1}(a) we examine the transition from a configuration in
series to a symmetrical parallel geometry, as done in Ref.
\cite{ghost}. The parameter $t$ is the coupling between dots,
$\varepsilon_1$ and $\varepsilon_2$ the energies of the quantum
dots, and $V^{\alpha}_{j}$ the coupling between the dot $j$-th and
the lead $\alpha=L,R$. We put $V_{1}^{L}=V_{2}^{R}\equiv
V_{0} $ and $V_{2}^{L}=V_{1}^{R}\equiv \eta V_{0}$ ($0\leq \eta \leq 1$%
), so that $\eta =0$ for the molecule connected in series, and
$\eta =1$ for the symmetric parallel configuration. For $\eta\neq
0,1$, the conductance displays Fano and Breit-Wigner resonances
associated to the molecular states, and as $\eta$ goes from $0$ to
$1$ there is a progressive reduction of tunneling through the
bonding state, which stop transmitting when $\eta=1$ \cite{ghost}.

In Fig. \ref{Fig1}(b) we assume
two quantum dots with energies $\varepsilon_1=\varepsilon_0+\Delta$ and $%
\varepsilon_2=\varepsilon_0-\Delta$ side-attached to the same site
of a quantum wire, with equal coupling constants, $V_0$. The
conductance in this case exhibits a Breit-Wigner and two Fano
resonance around $\varepsilon_0-\Delta$ and
$\varepsilon_0+\Delta$, respectively. When the energies of the
quantum dot levels are equal, that is $\Delta=0$, one of the
resonances disappears for the arising of a BIC \cite{malyshev}.
Each of the systems can be represented by a scattering matrix $S$,
which is expressed in terms of the retarded Green's function
$\mathbf{G}^{r}$ by means of the Fisher-Lee relation \cite{fisher}
\begin{equation}
S_{\alpha,\beta}(\varepsilon)=-\delta_{\alpha\beta}+i\hbar\sqrt{v_\alpha
v_\beta} G^r_{\alpha,\beta}(\varepsilon),  \label{eq-7}
\end{equation}
where $\alpha,\beta=L,R$, $v_{\alpha}=d\varepsilon/dk$ is the
electronic velocity in the lead $\alpha$. For the double molecule
of Fig. \ref{Fig1}(a), if $\varepsilon_1=\varepsilon_2=0$, the
matrix elements of the scattering matrix are
\begin{eqnarray}
S_{LL} &=&S_{RR}=\frac{(t-\varepsilon )(t+\varepsilon )-4\Gamma
^{2}(\eta
^{2}-1)^{2}}{\Lambda }  \notag \\
S_{LR} &=&S_{RL}=\frac{-4i\Gamma \lbrack t(\eta ^{2}+1)-2\eta
\varepsilon ]}{\Lambda }  \label{smat}
\end{eqnarray}%
with
\begin{equation}
\Lambda =(t-\varepsilon -2i\Gamma (\eta -1)^{2})(t+\varepsilon
+2i\Gamma (\eta +1)^{2}),
\end{equation}%
where $\Gamma =V_{0}^{2}/2v$, $v$ being the hopping in the leads.
With the scattering matrix elements given in Eq. (\ref{smat}), the
Friedel phase $\theta _{F}$ can be obtained from Eq.
(\ref{friedelphase}), giving
\begin{eqnarray}
\theta _{F} &=&-\frac{1}{2}\arctan {\frac{4\Gamma (1+\eta
)^{2}(\varepsilon -t)}{-4\Gamma ^{2}(1+\eta )^{4}+(\varepsilon
-t)^{2}}}  \notag \\
&&-\frac{1}{2}\arctan {\frac{4\Gamma (-1+\eta )^{2}(\varepsilon +t)}{%
-4\Gamma ^{2}(-1+\eta )^{4}+(\varepsilon +t)^{2}}}  \label{fried1}
\end{eqnarray}%
Replacing Eq. (\ref{fried1}) in (\ref{fsr}) we get the density of
states
\begin{widetext}
\begin{equation}
\rho(\varepsilon)=\frac{4\Gamma\left[(1+\eta^2)(4\Gamma^2(-1+\eta^2)^2+\varepsilon^2)+
4\eta\varepsilon
t+(1+\eta^2)t^2\right]}{\pi[4\Gamma^2(1+\eta)^4+(\varepsilon-t)^2]
[4\Gamma^2(-1+\eta)^4+(\varepsilon+t)^2]}. \label{dos1}
\end{equation}
\end{widetext}
For the two side-attached quantum dots the scattering matrix
elements are
\begin{eqnarray}
S_{LL} &=&S_{RR}=\frac{-2i\varepsilon \Gamma }{\Lambda }  \notag \\
S_{LR} &=&S_{RL}\frac{\left( \varepsilon -\Delta \right) \left( \varepsilon
+\Delta \right) }{\Lambda }  \label{smat2}
\end{eqnarray}%
with $\Lambda =\left( \varepsilon -\Delta \right) \left(
\varepsilon +\Delta \right) +2i\varepsilon \Gamma $, with $\Gamma
=V_{0}^{2}/2v$. We have put $\varepsilon _{0}=0$. The above $S$%
-matrix elements lead to the following expression for $\theta _{F}$
\begin{equation}
\theta _{F}=\frac{1}{2i}\ln \left[ -\frac{\left( \varepsilon
-\Delta \right) \left( \varepsilon +\Delta \right) -2i\varepsilon
\Gamma }{\left( \varepsilon -\Delta \right) \left( \varepsilon
+\Delta \right) +2i\varepsilon \Gamma }\right], \label{fried2}
\end{equation}
and the density of states is
\begin{equation}
\rho \left( \varepsilon \right) =\frac{1}{\pi }\frac{2\left(
\varepsilon ^{2}+\Delta ^{2}\right) \Gamma }{\left[ \left(
\varepsilon -\Delta \right) \left( \varepsilon +\Delta \right)
\right] ^{2}+4\varepsilon ^{2}\Gamma ^{2}}.
\\
\label{dos2}
\end{equation}%
\begin{figure}[hb]
\centering
\includegraphics[scale=0.2,angle=-90]{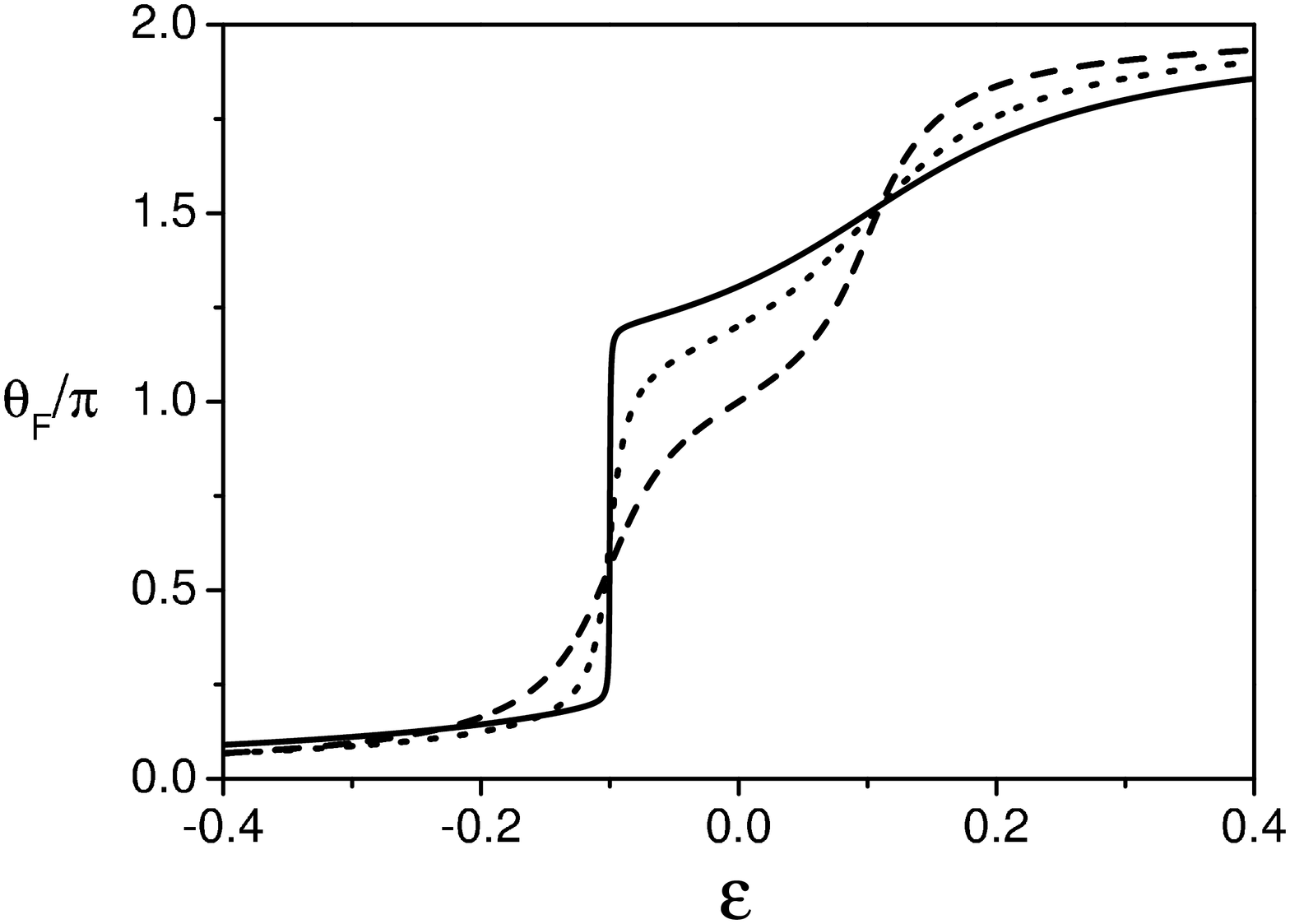} %
\centering
\includegraphics[scale=0.2,angle=-90]{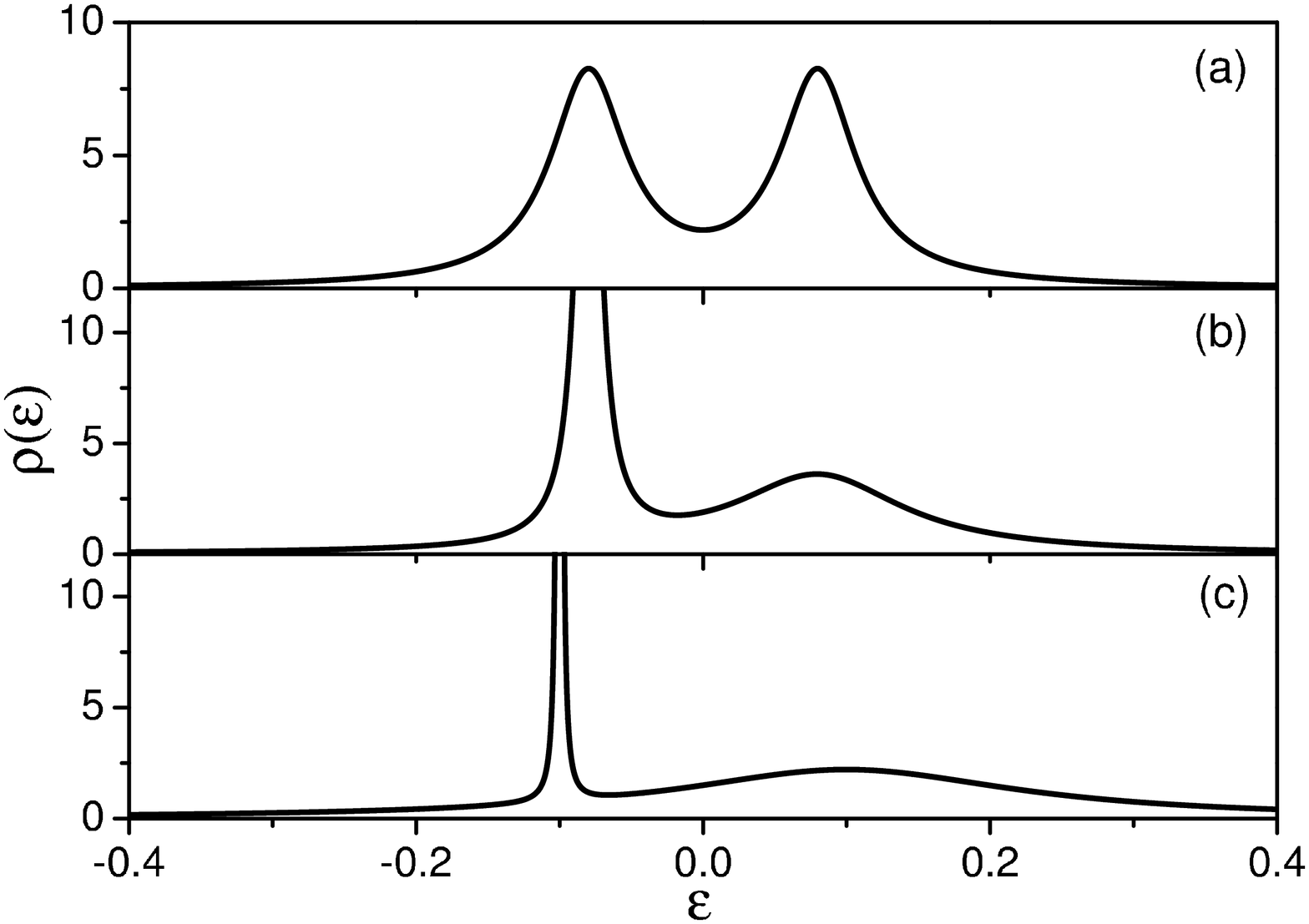}
\caption{Upper panel: Friedel phase versus energy for the double
molecule, for $t=0.1$ and
different values of $\protect\eta$: $\protect\eta=0$ (dash line), $%
\protect\eta=0.5$ (dotted line) and $\protect\eta=0.9$ (solid
line). Lower panel: Density of states for a) $\protect\eta =0$
(configuration in series), b) $\protect\eta =0.5$ and c) $\protect\eta %
=0.9$. } \label{Fig2}
\end{figure}
Fig. \ref{Fig2} (upper panel) shows the Friedel phase for the
double quantum dot molecule for three different values of $\eta$.
For $\eta=0$ (configuration in series) $\theta_F$ increases
smoothly in $\pi$ around each of the resonances. When $\eta$ gets
larger, that is, we go from a configuration in series to a
parallel one, the Friedel phase changes more steeply around the
energy of the bonding state, $\varepsilon_-=-t$. In the limit
$\eta \rightarrow 1$ (symmetric parallel configuration), where the
BIC takes place, $\theta_F$ increases abruptly, becoming
discontinuous. The lower panel shows the corresponding density of
states. This in general is a superposition of two Lorentzians at
the molecular energies, with broadenings $\Gamma _{-}=\Gamma
(-1+\eta )^{2}$ and $\Gamma _{+}=\Gamma (1+\eta )^{2}$. For the
configuration in series, the two states have the same width. When
$\eta $ approaches to $1$, one of the peaks becomes progressively
narrower than the other, as shown in figures (b) and (c). In the
limit $\eta \rightarrow 1$, $\Gamma _{-}\rightarrow 0$ and the DOS
at the bonding energy is a Dirac delta.

\begin{figure}[h]
\centering
\includegraphics[scale=0.22,angle=-90]{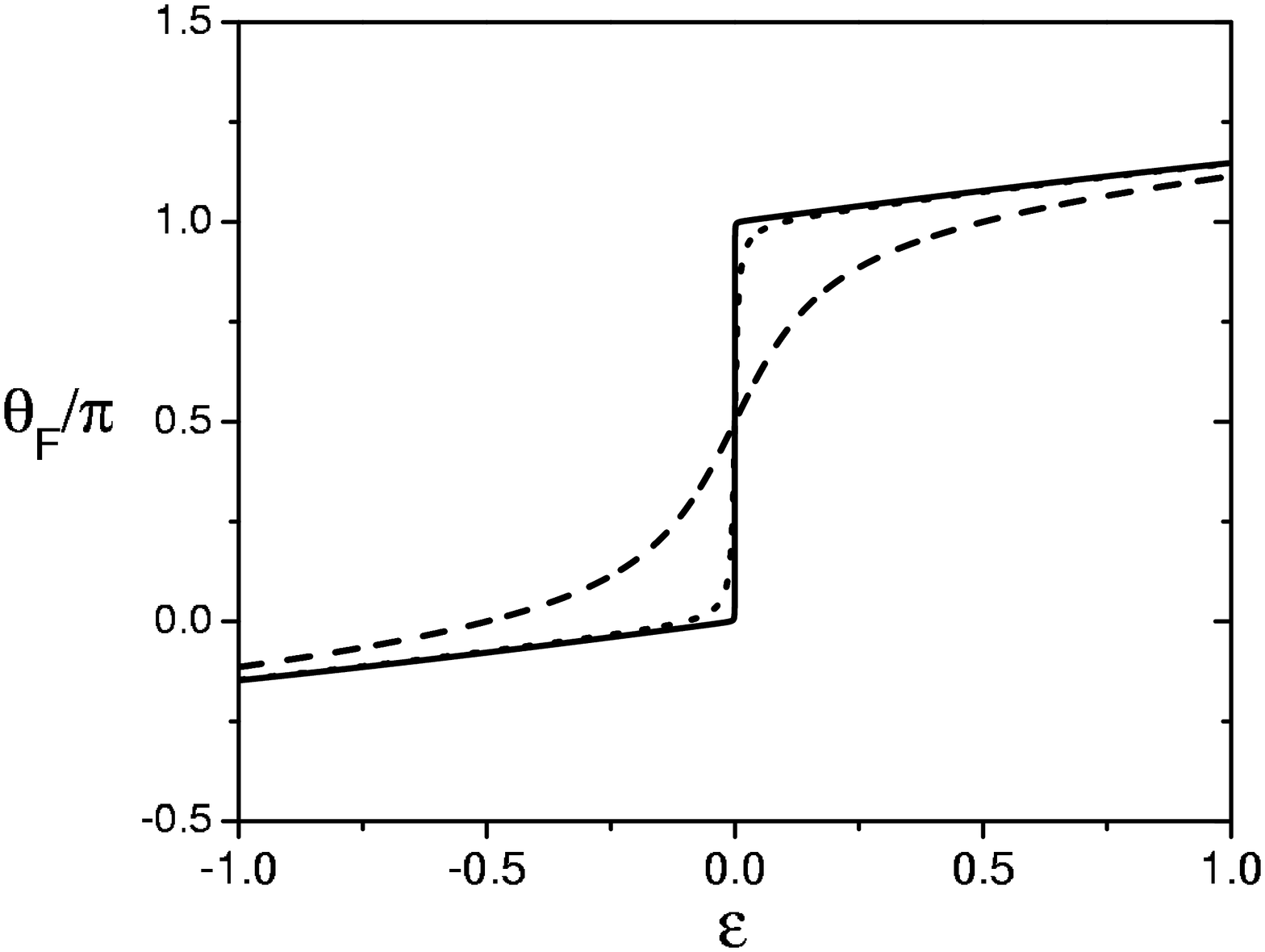}\newline
\centering
\includegraphics[scale=0.2,angle=-90]{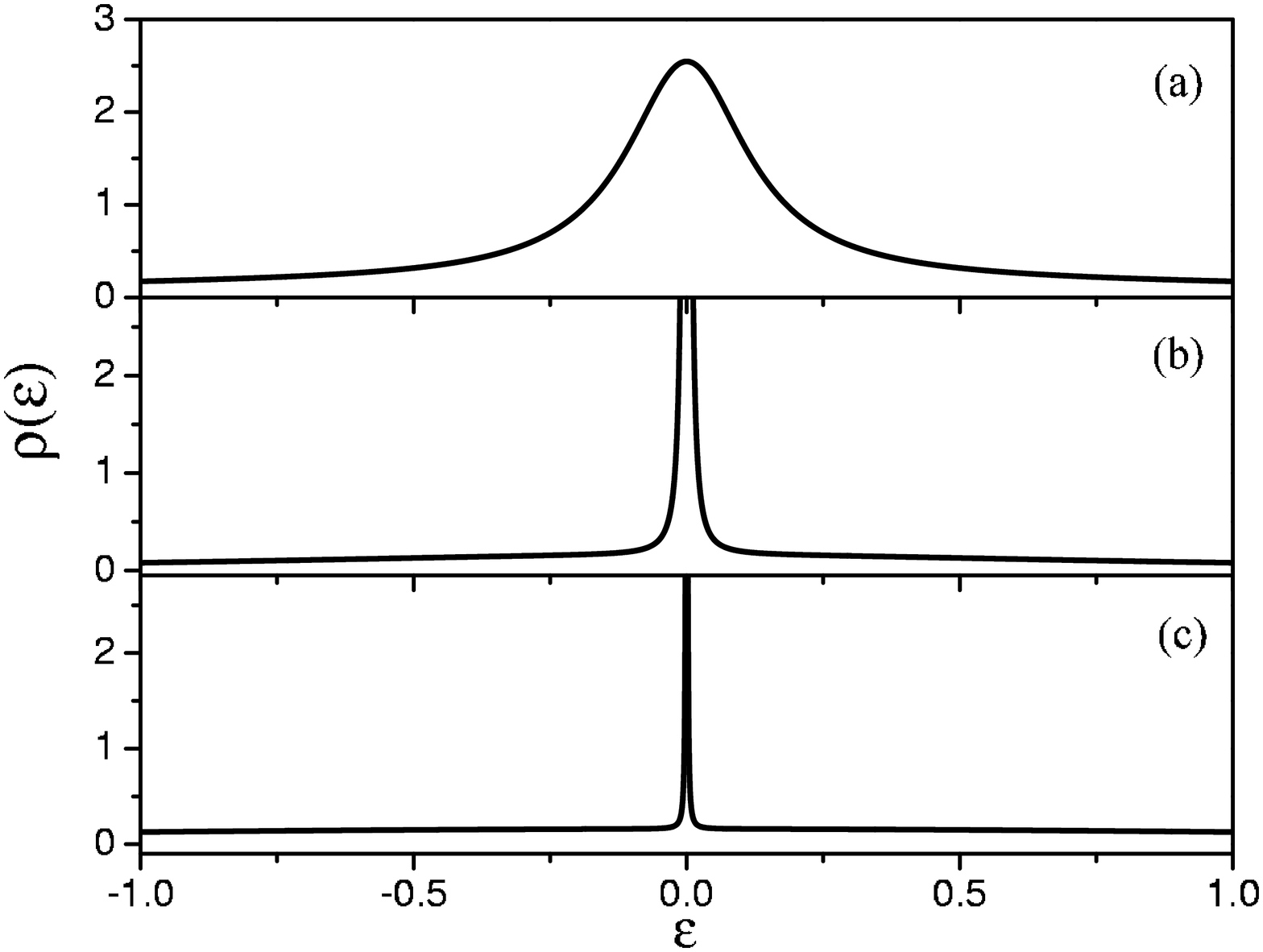}
\caption{Upper panel: Friedel phase for the side-coupled dots for
different values of $\Delta$: $\Delta=0.5$ (dash line),
$\Delta=0.1$ (dotted line), $\Delta=0.01$ (solid line). Lower
panel: Density of states for a) $\Delta =0.5$, b) $\Delta =0.1$,
c) $\Delta =0.01$  } \label{Fig3}
\end{figure}

The behavior of the Friedel phase in presence of a BIC is
analogous for the side-coupled quantum dots, as shown in Fig.
\ref{Fig3} (upper panel). For $\Delta$ of the order of $\Gamma$,
$\theta_F$ changes smoothly around $\varepsilon_0$ (dash line),
but it suffers an abrupt jump when $\Delta$ approaches to $0$,
when one of the resonances becomes a BIC. Fig. \ref{Fig3} (lower
panel) shows the corresponding density of states. As shown in Ref.
\cite{malyshev}, when $\Delta \ll \Gamma$, as in (b) and (c), the DOS
can be approached by a sum of two Lorentzians of widths $\Gamma
_{+}=2\Gamma $ and $\Gamma _{-}=\Delta ^{2}/2\Gamma$ centered at
$\varepsilon_0$.

In summary, for any of the considered systems, the density of
states when $\Gamma_-\rightarrow 0$ can be written as a
superposition of a Lorentzian and a Dirac delta function
\begin{equation}
\rho \left( \varepsilon \right) =\frac{1}{\pi }\frac{2\Gamma }{\left(
\varepsilon -\varepsilon _{+}\right) ^{2}+4\Gamma ^{2}}+\delta
\left(\varepsilon -\varepsilon_-\right),
\end{equation}
where $\varepsilon_\pm=\pm t$ for the parallel-coupled double quantum dot molecule, and $%
\varepsilon_-=\varepsilon_+=\varepsilon_0$ for the side-coupled
quantum dots. Integrating the above equation in the energy
$\varepsilon$ we obtain the Friedel phase
\begin{equation}
\theta _{F}\left( \varepsilon \right)=\arctan \left( \frac{\varepsilon
-\varepsilon _{+}}{2\Gamma }\right) +\frac{\pi }{2}+\pi \Theta \left(
\varepsilon -\varepsilon _{-}\right).
\end{equation}
This equation describes correctly the abrupt jump exhibited by
$\theta_F$ in Figs. \ref{Fig2} and \ref{Fig3}, in the presence of
a BIC. The integration of Eq. (\ref{fsr}) allows to obtain the
charge $N_d$ in the quantum dots,
\begin{eqnarray}
N_{d}&=&\frac{e\theta _{F}\left( \mu \right) }{\pi } \nonumber \\
&=& \frac{e}{\pi }\arctan \left( \frac{\mu -\varepsilon
_{+}}{2\Gamma }\right) +e\Theta \left( \mu -\varepsilon
_{-}\right) +\frac{e}{2}, \label{charge}
\end{eqnarray}%
where $\mu $ is the Fermi energy. Thus, if $\varepsilon _{-}$
falls below the Fermi energy, the charge in the system formed by
the two dots increases abruptly in an unit. This discontinuity in
the charge in presence of a BIC should have an effect in the
conductance when the electron-electron interactions are taken into
account.

\begin{figure}[h]
\includegraphics[scale=0.3,angle=-90]{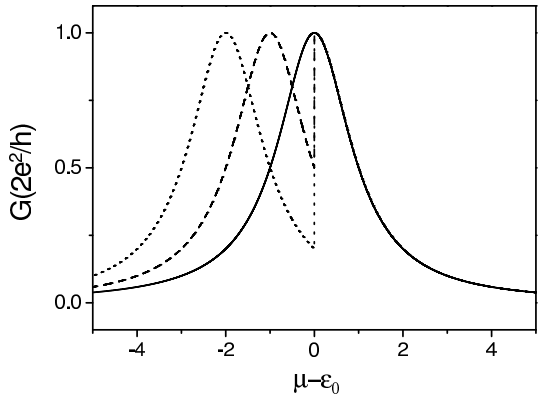}\newline
\caption{Conductance as a function of the Fermi energy, for a system of
interacting quantum dots in the presence of BICs, with $%
U=0$ (solid line), $U=2$ (dash line) and $U=4$ (dotted line).}
\label{conduct}
\end{figure}

Let us consider a generic system exhibiting BICs, which could be
two quantum dots in parallel, and let us treat the interaction by
the mean field approximation of the Anderson model. In this
approximation, the energy levels of the quantum dots are
renormalized, that is, if the energies of the quantum dots are
$\varepsilon _{1}=\varepsilon _{2}=\varepsilon _{0}$, the
interaction moves them to $\widetilde{\varepsilon
}_{0}=\varepsilon _{0}+UN_{d}/2$. Then, if $\varepsilon _{0}$
falls below the Fermi energy, the jump in $e$ produced in $N_{d}$
according to Eq. (\ref{charge}) will shift the renormalized energy
$\widetilde{\varepsilon }_{0}$ in $-U/2$, with consequences in the
conductance. In fact, the conductance can be written in terms of
the renormalized energy as
\begin{equation}
G=\frac{2e^{2}}{h}\frac{4\Gamma ^{2}}{\left( \mu -\widetilde{\varepsilon }%
_{0}\right) ^{2}+4\Gamma ^{2}}.
\end{equation}
If the quantum dots are initially empty, and if $\mu -\varepsilon
_{0}\rightarrow 0^{-}$, then
\begin{equation}
G\rightarrow \frac{2e^{2}}{h},
\end{equation}
but if $\mu -\varepsilon _{0}\rightarrow 0^{+}$
\begin{equation}
G\rightarrow \frac{2e^{2}}{h}\frac{%
4\Gamma ^{2}}{\frac{U^{2}}{4}+4\Gamma ^{2}}.
\end{equation}
Thus, $G$ suffers a discontinuity at $\varepsilon _{0}=\mu $,
as illustrated in Fig. \ref{conduct}.
In the absence of interaction the conductance profile is a
Lorentzian (solid line). If $U\neq 0$, when the Fermi energy
crosses the BIC (in $\mu =\varepsilon _{0}$) the resonance shifts
in $-U/2$, as shown for $U=2$ and $U=4$ (dash y dotted line,
respectively). In other words, the sudden filling of a BIC, as
described by Eq. (\ref{charge}), shifts the energy of the
transmitting state, with the subsequent arising of a discontinuity
in the conductance. This rough description allows to gain more
insight of the results shown by B\"{u}sser \emph{et al.} in
multilevel quantum dots in Kondo regime \cite{busser}. These
authors found similar discontinuities in the conductance as the
gate voltage was varied, when a localized state crossed the Fermi
energy. These localized states are a many-body version of the
bound states in the continuum discussed here. This effect is
analogous to that of a charge sensing described in
\cite{berkovits}, but in the present scheme, the sensed charge is
that of a BIC inside the proper system.

In summary, we have studied the behavior of the Friedel phase in
quantum dots systems when BICs are produced. This phase exhibits
an abrupt change when these states arise, effect with consequences
in the charge of the system, which jumps in a unity when the
energy of the BIC drops below the Fermi energy.

\section*{Acknowledgments}

M. L. L. de G. acknowledges financial support from FONDECYT, under grant
1040385, and from Milenio ICM P02-049-F, and P. O. thanks financial support
from CONICYT/Programa Bicentenario de Ciencia y Tecnolog\'{\i}a (CENAVA,
grant ACT27).

\end{document}